\documentclass[aps,prb,a4paper,10pt,twocolumn,showpacs,floatfix,bibnotes,superscriptaddress,preprintnumbers,longbibliography]{revtex4-2}
\usepackage{graphicx} % Include figure files
\usepackage{dcolumn}  % Align table columns on decimal point
\usepackage{bm}       % bold math

\usepackage{soul,multirow}
\usepackage{color}
\usepackage[colorlinks,bookmarks=false,citecolor=darkblue,linkcolor=red,urlcolor=blue]{hyperref}
\usepackage[utf8]{inputenc}
\DeclareUnicodeCharacter{2062}{\,}
\DeclareUnicodeCharacter{03BC}{\,}
\DeclareUnicodeCharacter{1D445}{\,}
\DeclareUnicodeCharacter{1D40E}{\,}
\DeclareUnicodeCharacter{2009}{\,}
\usepackage[charter,greekuppercase=italicized]{mathdesign}

\usepackage{microtype}
\usepackage{afterpage}
\graphicspath{{../}{./}{figures/}}

\definecolor{darkred}{rgb}{0.7,0.0,0.0}

\definecolor{darkblue}{rgb}{0,0.02,0.45}

\definecolor{darkgreen}{rgb}{0.02,0.45,0.0}

\definecolor{violet}{rgb}{0.8,0.2,0.6}

\begin{document}

%\preprint{APS/123-QED}
\title{Proximate quantum spin liquid state in the frustrated HoInCu$_4$ metal}
\author{I. Ishant}
\affiliation{Department of Physics, Shiv Nadar Institution of Eminence, Gautam Buddha Nagar, Uttar Pradesh 201314, India}

\author{T. Shiroka}
\affiliation{{PSI Center for Neutron and Muon Sciences CNM, 5232 Villigen PSI, Switzerland}}
\affiliation{{Laboratorium f\"ur Festk\"orperphysik, ETH Z\"urich, 8093 Z\"urich, Switzerland}}

\author{O. Stockert}
\affiliation{Max Planck Institute for Chemical Physics of Solids, Dresden, Germany}

\author{V. Fritsch}
\affiliation{Experimental Physics VI, Center for Electronic Correlations and Magnetism, University of Augsburg, 86159 Augsburg, Germany}

\author{M. Majumder}
\email{mayukh.majumder@snu.edu.in}
\affiliation{Department of Physics, Shiv Nadar Institution of Eminence, Gautam Buddha Nagar, Uttar Pradesh 201314, India}

\date{\today}

\begin{abstract}

We conducted a comprehensive and comparative muon-spin relaxation and rotation ({\textmu}SR) investigation on two fcc-lattice metallic compounds, HoCdCu$_4$ ($T_\mathrm{N}\approx 8$\,K) and HoInCu$_4$ ($T_\mathrm{N}\approx 0.76$\,K), to elucidate the nature of their magnetic ground states and the role of frustration in stabilizing them. Our {\textmu}SR results reveal that, in contrast to HoCdCu$_4$, strong magnetic frustration
exist in HoInCu$_4$. %,consistent with previous findings.
Notably, in HoInCu$_{4}$, only 30\% of the Ho-moments participate in the static magnetic ordering below $T_\mathrm{N}$, while the remaining 70\% of the Ho-moments exhibit dynamic correlations and persistent spin dynamics down to 0.3\,K, resembling a quantum spin-liquid (QSL) behavior. By contrast, in HoCdCu$_{4}$, all the Ho-moments contribute to the magnetic order below $T_\mathrm{N}$.
Furthermore, in HoInCu$_{4}$, the temperature dependence of the relaxation rate indicates the presence of quantum critical fluctuations in the paramagnetic state near $T_\mathrm{N}$, suggesting the proximity to a quantum critical point (QCP).
These observations suggest that the ground state of HoInCu$_{4}$ is a proximate quantum spin liquid (PQSL), a state that has not been reported before in frustrated metallic systems. Our {\textmu}SR findings are further corroborated by recent inelastic neutron results on HoInCu$_4$, which show similarities to other insulating PQSL candidates, thus reinforcing our conclusions.

\end{abstract}

\maketitle

Heavy fermion (HF) systems have long served as a rich platform for exploring exotic matter states, including non-Fermi liquid behavior, quantum criticality, unconventional superconductivity, topological insulators, Weyl semimetals, hidden order, and charge density waves, among others~\cite{Morosan2012, Ramakrishnan2008, Senthil2003, Senthil2004, Qimiao2001, Hirsch2015, Xiao-Liang2011, Yan2017, Hassinger2008, Isakov2006}. Despite the diversity of this landscape, the interplay between magnetic frustration and other interactions in the HF systems remains largely unexplored, offering new horizons for discovery in condensed matter physics. Frustration arises when competing exchange interactions among magnetic moments cannot be simultaneously satisfied, often dictated by the particular lattice geometry.

The presence of frustration can suppress long-range magnetic ordering (LRO),
thereby stabilizing highly entangled quantum spin liquid (QSL) states. Theoretical models propose a global phase diagram~\cite{Si2010} for HF systems incorporating frustration. They suggest the emergence of different exotic states, such as QSL and various quantum critical points (QCP), where LRO is tuned to zero temperature, leading to non-Fermi liquid (NFL) behavior. However, experimental evidence of such frustration-driven spin liquid states in metallic HF systems is scarce due to several challenges: (i) the long-range nature of RKKY interaction stabilizes an LRO, (ii) Kondo interactions tends to screen magnetic moments, thus diminishing frustration, (iii) QSL was originally proposed for local moments or Mott insulators, whereas HF systems typically involve itinerant moments. Moreover, in the Fermi-liquid state, both the susceptibility ($\chi$) and the heat capacity over temperature ($C/T$) follow a temperature-independent behavior. In a simplistic  situation, also a QSL state may exhibit the same behavior due to the presence of spinon Fermi surface. Clearly, the  stabilization of a QSL, as well as its unambiguous experimental demonstration is rather difficult in metals.  

CeRhSn~\cite{Tokiwa2015}, Pr$_2$Ir$_2$O$_7$~\cite{Nakatsuji2006}, and CeIrSn~\cite{Shimura2021} were claimed to be QSL candidate materials. Moreover, a chemical pressure induced QSL state has been proposed for CePd$_{1-x}$Ni$_x$Al~\cite{Ishant2024} and CeRh$_{1-x}$Pd$_x$Sn~\cite{Tripathi2022}. Remarkably, CePdAl under hydrostatic pressure exhibits a critical SL state with quantum critical fluctuations~\cite{Majumder2022}. Recently, evidence of a rare occurrence of field-induced critical spin liquid (CSL) has also been reported~\cite{Ishant2025}. Interestingly, in addition to the QSL or CSL states induced by frustration, a distinct state induced by frustration has been observed: the ``proximate quantum spin liquid'' (PQSL). Experimental evidence of the PQSL state has only been reported in  insulators, such as K$_2$Ni$_2$(SO$_4$)$_3$~\cite{Ivica2021,Gonzalez2024}, KYbSe$_2$~\cite{Scheie2024}, and Cs$_2$CuCl$_4$~\cite{Coldea2003}, Ba$_3$CoSb$_2$O$_9$~\cite{Zhou2012}, BaCo$_2$(AsO$_4$)$_2$~\cite{Zhang2022,Halloran2022} and Na$_2$Co$_2$TeO$_6$~\cite{Jiao2024} and $\alpha$-RuCl$_3$\cite{Banerjee2016}. The experimental signatures of such state are expected to be the following: 

(i) A low magnetic ordering temperature $T_\mathrm{N}$, indicative of closeness to a QCP, along with a low spin value, that enhances quantum fluctuations, as well as strong magnetic frustration, signaling the proximity to a QSL state.

(ii) Coexistence of both static- and dynamic magnetic correlations below $T_\mathrm{N}$, reflecting a partial magnetic volume fraction alongside persistent spin dynamics.

(iii) The presence of quantum critical fluctuations above $T_\mathrm{N}$, consistent with the proximity to a QCP. 

As PQSL states have not yet been observed in any frustrated metallic system, we are motivated to explore the possibility of realizing them in a frustrated metallic compound.

To this end, here we focus on the fcc-lattice compound HoInCu$_4$. A recent investigation has shown that it adopts a type-III AFM ordering below $T_\mathrm{N} \approx 0.76$\,K~\cite{Stockert2020}, with only half of the Ho$^{3+}$ moments taking part in the ordered state, as inferred from the entropy release associated with the ordering, corroborated by neutron diffraction data. The combination of strong magnetic frustration and a low effective spin value ($S_{eff} = 1$), along with a low ordering temperature, positions HoInCu$_4$ as an ideal candidate
for exhibiting a proximate quantum spin liquid (PQSL) in its ground state.

To explore the exotic ground state of HoInCu$_4$, a microscopic tool that can probe
its static and dynamic properties is of the utmost importance.
In our case, this coincides with the muon-spin relaxation/rotation ({\textmu}SR) technique,
which can also be used to identify the magnetic phase separation expected to occur in a PQSL.
The {\textmu}SR experiments on HoCdCu$_4$ and HoInCu$_4$ were performed on the GPS ($T_\mathrm{min} = 1.5$\,K) and VMS ($T_\mathrm{min} = 0.3$\,K) spectrometers respectively of the Swiss Muon Source (S{\textmu}S) at the Paul Scherrer Institute, Switzerland. For a comparison, we also employed {\textmu}SR on the non-frustrated compound HoCdCu$_4$, which exhibits a type-II AFM ordering at 8\,K. 
Despite sharing the same fcc lattice, HoCdCu$_4$ has a higher density
of states at the Fermi level, which promotes additional RKKY exchange interactions and therefore 
the next-nearest-neighbor exchanges becomes non-negligible which further promotes the  magnetic ordering~\cite{Stockert2020} by reducing the frustration.
Consequently, by comparing the nature of the static and dynamic
properties of HoInCu$_4$ and HoCdCu$_4$, we can unambiguously identify
the role of frustration in governing the ground state of HoInCu$_4$. 

%%%%%%%%%%%%%%%%%%%%%%%%%%%%%%%%%%%%%%%%%%%%%%%%%%%%%%%%%%%%%%%%
\begin{figure}
{\centering {\includegraphics[width=8cm]{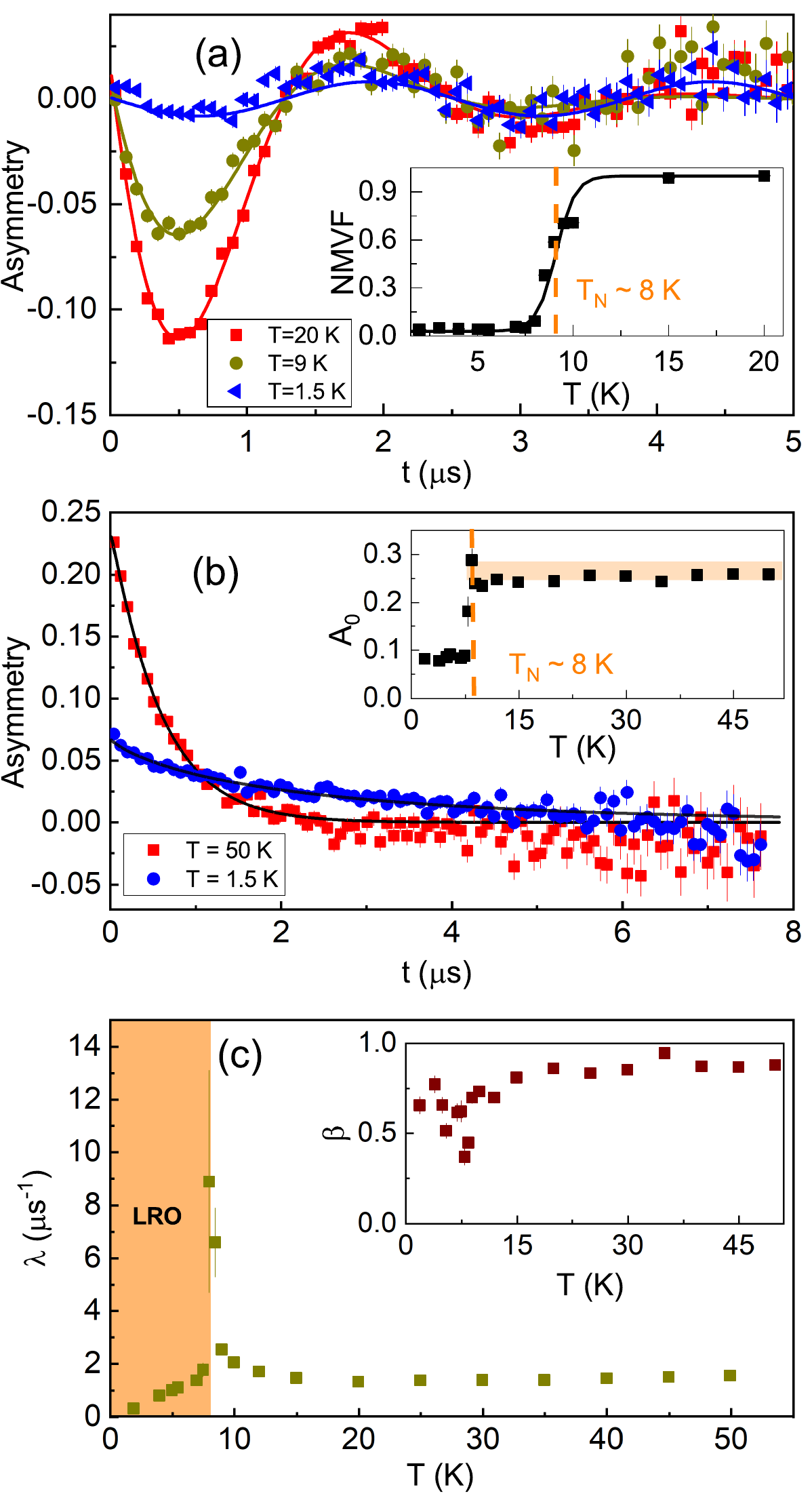}}\par}\caption{\label{fig:asym_vs_time} (a) wTF-{\textmu}SR asymmetry at different temperatures for HoCdCu$_4$. Inset: Non-magnetic volume fraction (NMVF) as a function of temperature,
as determined from the wTF measurements. Inset: Non-magnetic volume fraction (NMVF) as
a function of temperature, as determined from the wTF measurements.
The solid lines represent a fit using the sigmoidal function. (b) Asymmetry vs.\ time for HoCdCu$_4$ at selected temperatures. Inset: Initial asymmetry as a function of temperature for HoCdCu$_4$. (c) The temperature dependence of $\lambda$ as a
function of temperature peaks at $T_\mathrm{N} = 8$\,K. Inset:
$\beta$ as a function of temperature. LRO stands for long-range magnetic ordering.}
\end{figure}    
%%%%%%%%%%%%%%%%%%%%%%%%%%%%%%%%%%%%%%%%%%%%%%%%%%%%%%%%%%%%%%%%%%%%

First, we discuss the {\textmu}SR results of HoCdCu$_4$. Weak transverse field (wTF) {\textmu}SR experiments provide information about the magnetic volume fraction across the magnetic ordering temperature. The wTF spectrum of HoCdCu$_4$
can be well fitted by an exponentially decaying cosine function as shown in Fig.~\ref{fig:asym_vs_time}(a): 
\begin{equation}
\label{eq:magn_frac}
A(t) = A_{0} \cos (\omega t + \phi_\mathrm{TF})e^{-\lambda t}.
\end{equation}
Here, $A(t)$ is the muon spin polarization function. % under wTF. 
$A_{0}$, $\omega$, $\phi_\mathrm{TF}$, and $\lambda$ are the
asymmetry, the muon Larmor frequency, the initial phase, and the relaxation rate resulting from the applied field, respectively.
As seen in the inset of Fig.~\ref{fig:asym_vs_time}(a), the temperature dependence 
of the normalized asymmetry $A_{0}$ was fitted by the empirical
sigmoidal function $A(T) = A_a + \frac{A_b-A_a}{1+ \exp\frac{T-T_\mathrm{N}}{\Delta T}}$, where $T_\mathrm{N}$ is the ordering temperature, $\Delta T$ is the transition width, while $A_a$ and $A_b$ are the asymmetry above and below $T_\mathrm{N}$. As shown in the inset, the fit confirms that the ordering
temperature of HoCdCu$_4$ is $\approx 8$\,K, a result consistent
with those of magnetization and heat-capacity measurements.
More importantly, since $A_b$ is nearly zero, this indicates
that all the Ho$^{3+}$ moments are involved in the LRO.

%%%%%%%%%%%%%%%%%%%%%%%%%%%%%%%%%%%%%%%%%%%%%%%%%%%%%%%%%%%%%%%%
\begin{figure*}
{\centering {\includegraphics[width=18cm]{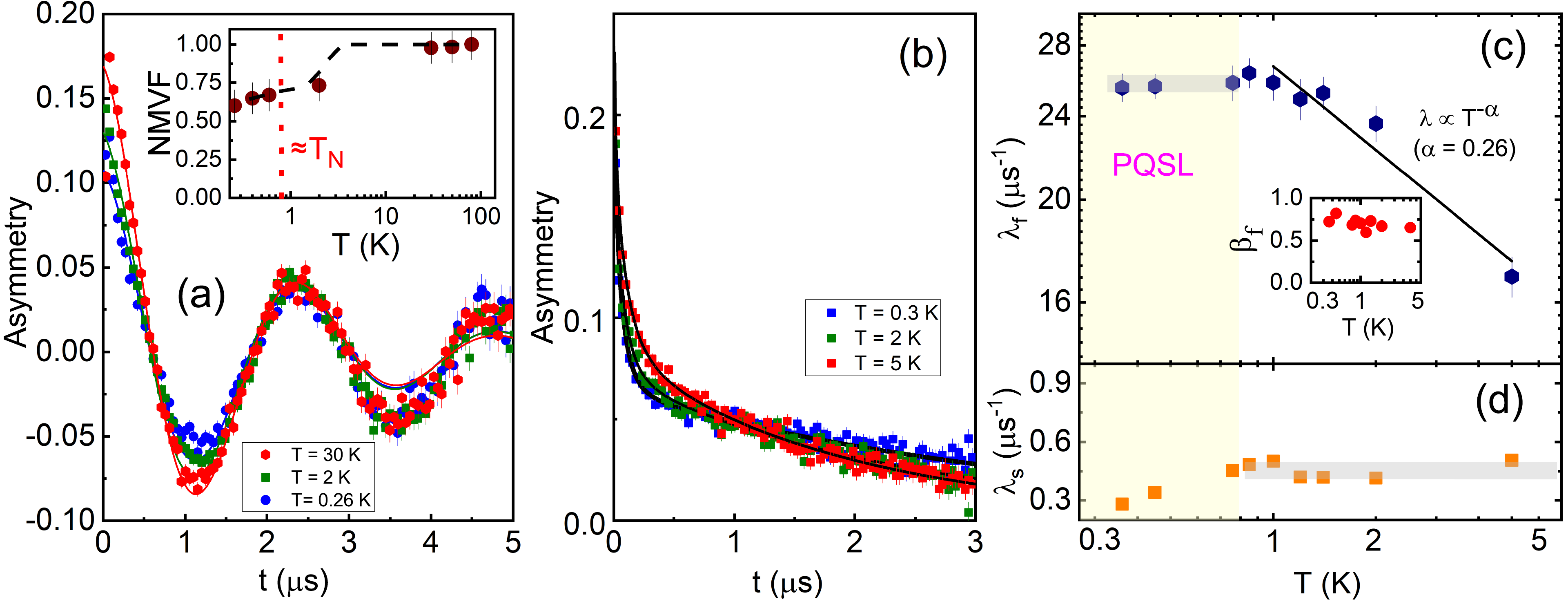}}\par}\caption{\label{fig:zf_musr} (a) wTF-{\textmu}SR asymmetry at different temperatures for HoInCu$_4$. Inset: Non-magnetic volume fraction (NMVF) as a function of temperature,
as determined from the wTF measurements. The dashed line is a
guide to the eyes. (b) Zero-field {\textmu}SR asymmetry of HoInCu$_4$ at different temperatures.
Solid lines represents two-exponential fit functions --- see Eq.~(\ref{eq:two_exp_decay}).
(c) and (d) represent the temperature dependence of $\lambda_f$ and $\lambda_s$ respectively.
%as a function of temperature % THIS IS TOTALLY REDUNDANT.
 Inset of Fig. 1(c): $\beta_f$ as a function of temperature.
See text for details.}
\end{figure*}
%%%%%%%%%%%%%%%%%%%%%%%%%%%%%%%%%%%%%%%%%%%%%%%%%%%%%%%%%%%%%%%%%%%%

The ZF-µSR technique can be used to determine the nature of an
ordered state and the magnitude of the local field produced by
magnetic moments at muon sites within that state.
Moreover, ZF-{\textmu}SR also provides information about the nature
of the spin dynamics in the paramagnetic state.
In the latter case, % Avoid repeting "paramagnetic state"
for $T > T_\mathrm{N}$, the ZF-{\textmu}SR asymmetry of HoCdCu$_4$
can be fitted by
\begin{equation}
\label{eq:relax}
A(t) = A_{0}e^{-\lambda t^{\beta}},
\end{equation}
where $A_0$ is the initial asymmetry, corresponding to muons stopped
in the sample, and $\lambda$ is the relaxation rate. The temperature
dependence of $\lambda$ is shown in Fig.~\ref{fig:asym_vs_time}(c).
At high temperatures (compared to the ordering temperature), $\lambda$ is
independent of temperature, but below 15\,K, $\lambda$
starts to increase and it diverges at the ordering temperature, 
$T_\mathrm{N}\approx 8$\,K. The enhancement of $\lambda$ with lowering temperature towards $T_\mathrm{N}$, %the ordering temperature,
reflects the critical slowing down of spin-fluctuations due to the
development of electronic spin correlations.

Interestingly, no oscillations are observed in the ZF-{\textmu}SR
asymmetry below the ordering temperature $T_\mathrm{N}$. This
allows us to use Eq.~(\ref{eq:relax}) to fit the ZF-{\textmu}SR asymmetry
also below $T_\mathrm{N}$.
At the same time, the initial asymmetry value, $A_0$,
decreases when entering the ordered magnetic state [see the right
inset in Fig.~\ref{fig:asym_vs_time}(b)]. 
The drop in $A_0$ to one-third of its high-temperature value [from 0.25 to 0.083] 
is what is expected in a polycrystalline sample in its magnetically
ordered state. This indicates that, below $T_\mathrm{N}$, on average,
one-third of muon spins are aligned with the local magnetic field~\cite{Anand2017}. Note that such a behavior is commonly observed in long-range magnetically ordered systems, where muons experience a large static field along with very short relaxation times~\cite{Frandsen2020,Zappala2024, Wu1994}. 
As shown in Fig.~\ref{fig:asym_vs_time}(c), below the ordering temperature, $\lambda$ decreases as the temperature is lowered, suggesting
the development of strong electronic correlations. At the same
time, as shown in the inset, $\beta$ decreases from 1, indicating the
onset of inhomogeneous local magnetic fields. We also performed inverse Laplace transforms (ILT) on the raw zero-field {\textmu}SR time-domain data across the full temperature range, following the procedure reported in Ref.~\cite{Martin2016, Wang2021}. ILT provides a model-independent proof about the presence (or absence) of a magnetic phase separation. For HoCdCu$_4$, the representative ILT spectra shown in Fig.~S2~\cite{supply}, reveal a single dominant peak in the probability distribution, consistent with the absence of magnetic phase separation.

Overall, the wTF- and ZF-{\textmu}SR study of HoCdCu$_4$
confirms that all Ho moments participate in the LRO below 8\,K, 
and no signs of frustration are observed. In case of frustration,
$\lambda$ would remain high at temperatures much higher than the
ordering temperature. In HoCdCu$_4$, however, $\lambda$ decreases to a
constant value at temperatures less than twice $T_\mathrm{N}$.

We now turn our attention to HoInCu$_4$ %compound
and compare the relevant {\textmu}SR results with those obtained in the
non-frustrated HoInCu$_4$ case (see the above discussion). The temperature
dependence of the non-magnetic volume fraction, estimated from
the wTF-{\textmu}SR spectra fitted using Eq.~(\ref{eq:magn_frac}),
is shown in the inset of Fig.~\ref{fig:zf_musr}(a). Interestingly, as the temperature is lowered, the NMVF begins to decrease well above $T_\mathrm{N}$, with no significant reduction occurring below $T_\mathrm{N}$. This is in sharp contrast to HoCdCu$_4$. In HoInCu$_4$, a finite NMVF down to almost 0\,K, indicates a fraction of Ho$^{3+}$ moments that do not participate in the magnetic ordering. 

The ZF-{\textmu}SR asymmetry of HoInCu$_4$, throughout the temperature range (between 5\,-0.3\,K), was fitted using

\begin{equation}
\label{eq:two_exp_decay}
A(t) = A_0[fe^{-\lambda_ft^{\beta}} + (1-f)e^{-\lambda_st}],
\end{equation} 

where $A_{0}$ is the initial {\textmu}SR asymmetry, while
$\lambda_s$ and $\lambda_f$ are the slow and fast relaxation rates,
respectively. The fraction $f$ turns out to be $\approx70\%$. Further, ILT spectra at the lowest measured temperature supports the presence of two distinct relaxation rates~\cite{supply}. 

In the paramagnetic state ($T>T_\mathrm{N}$), as the temperature decreases towards $T_\mathrm{N}$, the obtained
$\lambda_f$ values increase, indicative of developing electronic correlations.
Notably, as shown in Fig.~\ref{fig:zf_musr}(c), the enhancement of $\lambda_f$
begins at temperatures significantly higher than (at least six times)
$T_\mathrm{N}$. Such early increase in $\lambda_f$ suggests the
onset of short-range correlations, a hallmark of frustrated
systems~\cite{Ishant2024}. The presence of frustration in HoInCu$_4$
is reinforced by the power-law behavior of $\lambda_f$, similar to that
occurring in the renowned frustrated compound CePdAl~\cite{Ishant2024,Majumder2022}.
The persistence of a long tail in the specific heat and the presence
of broad magnetic Bragg peaks above $T_\mathrm{N}$~\cite{Stockert2020} further
corroborate this scenario. By contrast, as illustrated
in Fig.~\ref{fig:asym_vs_time}(c), the enhancement of $\lambda$ in HoCdCu$_4$
begins just above $T_\mathrm{N}$. Collectively, these observations
suggest that frustration is pronounced in HoInCu$_4$, but negligible in HoCdCu$_4$.

On the contrary, $\lambda_s$ shows a temperature independent behavior, reflecting the significant dipolar contribution from the holmium and indium nuclei, whose magnetic moments are 4.17 and 5.54\,$\mu_\mathrm{N}$, respectively

In the ordered state ($T<T_\mathrm{N}$), neither oscillations
nor a reduction in the initial asymmetry are observed down to 0.3\,K. Interestingly, $\lambda_f$ saturates indicating persistent spin dynamics with substantial low-energy excitations. In contrast, $\lambda_s$ clearly reduces with lowering temperature below $T_\mathrm{N}$, which resembles the temperature dependence of relaxation rate for a system below ordering temperature. Hence, the ZF measurements indicates the presence of phase separation below $T_\mathrm{N}$.

\begin{figure}
{\centering {\includegraphics[width=8cm]{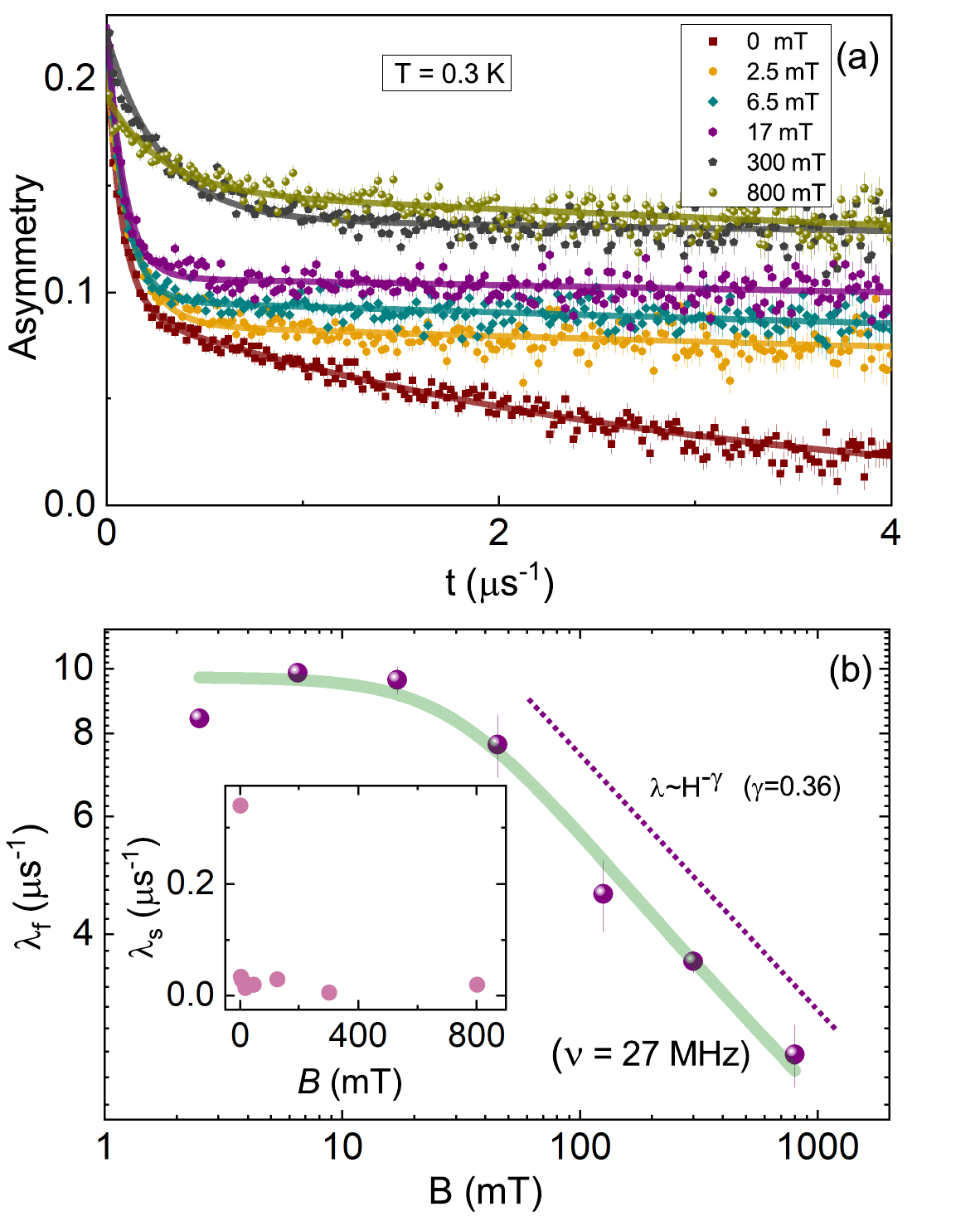}}\par}\caption{\label{fig:lf_musr}(a) Asymmetry as a function of time at different fields for HoInCu$_4$. (b) Fast relaxation rate $\lambda_f$ vs.\ longitudinal field at a base temperature (0.3\,K). Solid lines represent fits using Eq.~(\ref{eq:spin_dynamics}), while dashed lines are power-law fits, $\lambda \propto H^{-\gamma}$.
Inset: static relaxation rate as a function of field, measured at 0.3\,K.}
\end{figure}

{\textmu}SR measurements in a longitudinal field (LF) allowed us to
determine the nature of the electron-spin dynamics, i.e., whether
the correlations are static or dynamic. The LF-{\textmu}SR asymmetry, 
measured at different fields at 0.3\,K, is shown in
Fig.~\ref{fig:lf_musr}(a). We have seen that ZF-{\textmu}SR 
%and wTF-{\textmu}SR that there are
exhibits two relaxation channels, a fast- and a slow one
[see Eq.~(\ref{eq:two_exp_decay})]. By applying an LF of about 
2.5\,mT, the slow component saturates (i.e., decouples),
indicating that such component corresponds to 30\% of
Ho$^{3+}$ moments that exhibit static correlations.
By contrast, the faster component does not saturate even after
applying an LF of 0.8\,T, suggesting that the rest of the
Ho$^{3+}$  moments are dynamically correlated [see Fig.~\ref{fig:lf_musr}(a)].

The LF dependence of the fast- $\lambda_f$ and the slow $\lambda_s$
relaxation rates at 0.3\,K is shown % the "dependence" is shown.
in Fig.~\ref{fig:lf_musr}(b) and its inset, respectively.

To quantify the spin dynamics at low temperatures, we employed a fit
function that accounts for the spin fluctuations of 4$f$ electrons,
which couple with the implanted muons: 
\begin{equation}
\label{eq:spin_dynamics}
\lambda(H) = 2 \Delta^{2} \tau^{x'} \int_{0}^{\infty}\! t^{-x'} e^{-\nu t} \cos(2\pi \mu_{0} \gamma_{\mu}Ht)\,  \mathrm{d}t,
\end{equation}
where $t$ is time, $\tau$ is an early-time cut-off, $\Delta$ is the
width of the internal-field distribution, $\gamma_\mu$ is the muon's
gyromagnetic ratio, and $\nu$ is the fluctuation frequency of local moments.
Fitting the LF dependence of $\lambda$ yields a non-zero exponent $x' \neq 0$,
indicating that the spin-spin autocorrelation function deviates from a
simple exponential behavior, $C(t) = \exp(-\nu t)$. Instead, it follows
$C(t) = (\tau/t)^{x'} \exp(-\nu t)$, as shown with a solid line in
Fig.~\ref{fig:lf_musr}(b). The estimated fluctuation frequency $\nu$
for HoInCu$_4$ is approximately 27\,MHz. On the other hand,
in the high-LF regime, $\lambda$ follow a power-law behavior
$\lambda \propto H^{-\gamma}$, with $\gamma = 0.36$. Thus, the LF
measurements suggest the coexistence of static- and dynamic
correlations below $T_\mathrm{N}$.

The unequivocal presence of long-range magnetic ordering at $T_\mathrm{N}\sim 0.76$\,K
in HoInCu$_4$, as evidenced by specific heat and elastic neutron diffraction
measurements~\cite{Stockert2020}, is in contrast to the absence of characteristic
ordering signatures in the {\textmu}SR results. Below we discuss 
the possible reasons for the absence of oscillations (i.e., the spontaneous
muon precession due to local static fields in a magnetically ordered state)
in the {\textmu}SR spectra associated with 30\% of the static Ho$^{3+}$ moments. 

(i) One plausible explanation for this anomaly in the {\textmu}SR results, is the presence of a quasi-static ordered state, a phenomenon previously observed in systems beset by disorder~\cite{Ishant2024, Kenney2019, Kulbakov2021,Ziat2017}. The clear specific heat anomaly at $T_\mathrm{N}$ and the clear evidence of static long-range order from neutron scattering are incompatible with the presence of significant disorder. Moreover, the compound's %fixed
stoichiometry is not affected %been compromised
by chemical disorder, %which may induce disorder and
as revealed by a highly ordered crystal structure from XRD and
neutron scattering~\cite{Stockert2020}. 

(ii) Complex ordering phenomena, such as incom\-men\-su\-ra\-te- or multiple-$Q$
order, typically yield a broad distribution of local fields at the muon site, thereby potentially obscuring spontaneous muon precession, as observed in previous studies~\cite{Kenney2022, Kulbakov2021,Bouaziz2024}. However, considering that neutron scattering studies suggests a commensurate
order in HoInCu$_4$~\cite{Stockert2020}, we can exclude this possibility as well.

(iii) The absence of oscillation in the {\textmu}SR asymmetry could also be attributed to a vanishing local dipolar field at the muon site. It is important to mention that a clear signature of LRO has been observed in HoCdCu$_4$. As HoInCu$_4$ has the same structure, it is expected that the muon site will also be the same and, thus, muons will experience a static local field. For a more precise statement, we employed DFT-based calculations to determine the muon site in HoInCu$_4$ and subsequently estimated the dipolar field based on the spin arrangements inferred from neutron scattering measurements
(see the supplementary material for the details~\cite{supply}). Notably, our calculations reveal two equally probable muon sites in HoInCu$_4$, accompanied by a finite dipolar field. This finding enables us to conclusively rule out the cancellation of dipolar fields at the muon site as a viable explanation for the observed {\textmu}SR behavior. 

%%%%%%%%%%%%%%%%%%%%%%%%%%%%%%%%%%%%%%%%%%%%%%%%%%%%%%%%%%%%%%%%
%%%%%%%%%%%%%%%%%%%%%%%%%%%%%%%%%%%%%%%%%%%%%%%%%%%%%%%%%%%%%%%%%%%% 

(iv) Notably, fast fluctuations or motional narrowing condition may also contribute to the absence of oscillations in the {\textmu}SR asymmetry. The motional narrowing condition is ${\tau_\mathrm{c} \ll (\gamma_\mu B_\mathrm{loc}})^{-1}$ where $\tau_\mathrm{c}$ is the fluctuation time of $B_\mathrm{loc}$. To assess this scenario, we calculated the dipolar field at the muon site (see Table I in the supplementary material~\cite{supply}) by considering the type-III AFM spin structure inferred from neutron diffraction experiments. Subsequently, we could estimate the correlation time $\tau_\mathrm{c} = 2.43 \times 10^{-11}$\,s by means of the equation: 
\begin{equation}
\lambda_\mathrm{s} = {\gamma_\mu}^2 B_\mathrm{loc}^2 \tau_\mathrm{c}.
\end{equation}                    
The fulfillment of the motional narrowing condition is evident upon utilizing
the calculated values of $B_\mathrm{loc}$ and $\tau_\mathrm{c}$ (yielding 
$\gamma_\mu B_\mathrm{loc} \tau_\mathrm{c} \simeq 0.02 \ll 1$). This
unequivocally demonstrates that the moments responsible for the magnetic
ordering fluctuate very fast compared to the {\textmu}SR time window.
As a result no oscillation is visible in the {\textmu}SR asymmetry
below $T_\mathrm{N}$. A similar phenomenon has been observed in Tb$_2$Sn$_2$O$_7$~\cite{{Dalmas2006}},
whose $\tau_\mathrm{c}$ is found to be $8 \times 10^{-11}$\,s, i.e., 
of the same order of magnitude as in HoInCu$_4$. An analogous situation
has been observed in several pyrochlore systems~\cite{Xu2016,Mauws2018,Lago2005,Dunsiger2006}, as well as in other frustrated systems~\cite{Miao2024,Cai2019}. It is noteworthy that a clear magnetic Bragg peak
is seen in the neutron diffraction data of HoInCu$_4$. The difference
between the {\textmu}SR and neutron results is associated
with the difference in the energy scale of the two techniques. More specifically,
neutrons probe on a much shorter time scale ($\sim 10^{-14}$\,s), whereas
the time scale of {\textmu}SR is $\sim 10^{-11}$\,s.

From the above discussion, on the basis of the {\textmu}SR
results regarding both compounds, we have that:

(1) Unlike in HoCdCu$_4$, a signature of frustration is
evident from the significant increase of $\lambda$ at much higher
temperatures compared to $T_\mathrm{N}$ in HoInCu$_4$. 

(2) In contrast to HoCdCu$_4$, only 30\% of the Ho$^{3+}$ moments
participate in the static magnetic ordering below $T_\mathrm{N}$ in HoInCu$_4$.

(3) Interestingly, $\lambda_f$, associated with the dynamic muon
relaxation, is independent of temperature below $T_\mathrm{N}$.
This suggests a persistent spin dynamics, characteristics of QSL
materials, including a fluctuation frequency comparable to that
of other spin-liquid candidates~\cite{Baenitz2018,Li2016,Kundu2020}.
Thus, in HoInCu$_4$, the static and dynamic correlation coexist
below $T_\mathrm{N}$, whereby the dynamic part is consistent with
the experimental signatures of a QSL state. 

(4) It is important to mention that, in HoInCu$_4$, the relaxation
rate above $T_\mathrm{N}$ exhibits a power-law behavior, 
$\lambda = 1/T_{1\mu} \propto T \times \chi''(q,\omega)$ $\propto T^{-\alpha}$
with $\alpha = 0.26$. The exceptionally small value of $T_\mathrm{N}$
implies that the critical divergence of $T_\mathrm{N}$ is driven by an interplay of both thermal and quantum fluctuations. This observation is corroborated by similar measurements in CePdAl under pressure and Ni-doping, which exhibit a comparable magnitude of the critical exponent in the non-Fermi liquid regime.

The experimental observations mentioned above indicate that
HoInCu$_4$ satisfies all the criteria to have a PQSL ground state. Note
that, out of the three insulating PQSL candidates, {\textmu}SR
has been performed only in K$_2$Ni$_2$(SO$_4$)$_3$~\cite{Ivica2021} and Na$_2$Co$_2$TeO$_6$~\cite{Miao2024, Jiao2024}.
In both the cases, below $T_\mathrm{N}$, coexisting static and dynamic
correlations have been claimed, similar to the HoInCu$_4$ case. 

Now we compare the elastic and inelastic neutron results
regarding HoInCu$_4$~\cite{Stockert2020,Boraley2025} with
those related to the other insulating PQSL candidates.
Magnetic Bragg peaks associated with magnetic ordering have been
observed in all insulating compounds, alongside broad diffusive
excitations in the inelastic neutron scattering data~\cite{Kamiya2018, Halloran2022}.
As far as HoInCu$_4$ is concerned, a clear magnetic Bragg peak
was observed. Recent inelastic neutron scattering measurements
provide signatures of diffuse excitation, corroborating the
signature of PQSL found in the insulating compounds. Interestingly, recent theoretical studies based on the pseudo-fermion functional renormalization group (pf-FRG) have shown that around $J_2/J_1 \approx 0.5$ there is a manifold of degenerate low-energy states. These states give rise to an extended spin-liquid regime in the quantum model centered around the classical high degeneracy point~\cite{Revelli2019,Kiese2022}. Since, in the HoInCu$_4$ case, $J_2/J_1 \approx 0.45$ was obtained~\cite{Boraley2025}, this further supports the presence of PQSL state in HoInCu$_4$.

%%%%%%%%%%%%%%%%%%%%%%%%%%%%%%%%%%%%%%%%%%%%%%%%%%%%%%%%%%%%%%%%%%%%%%%%%%%%%%%%%%%%%%%%
\begin{figure}
\includegraphics[width=10cm]{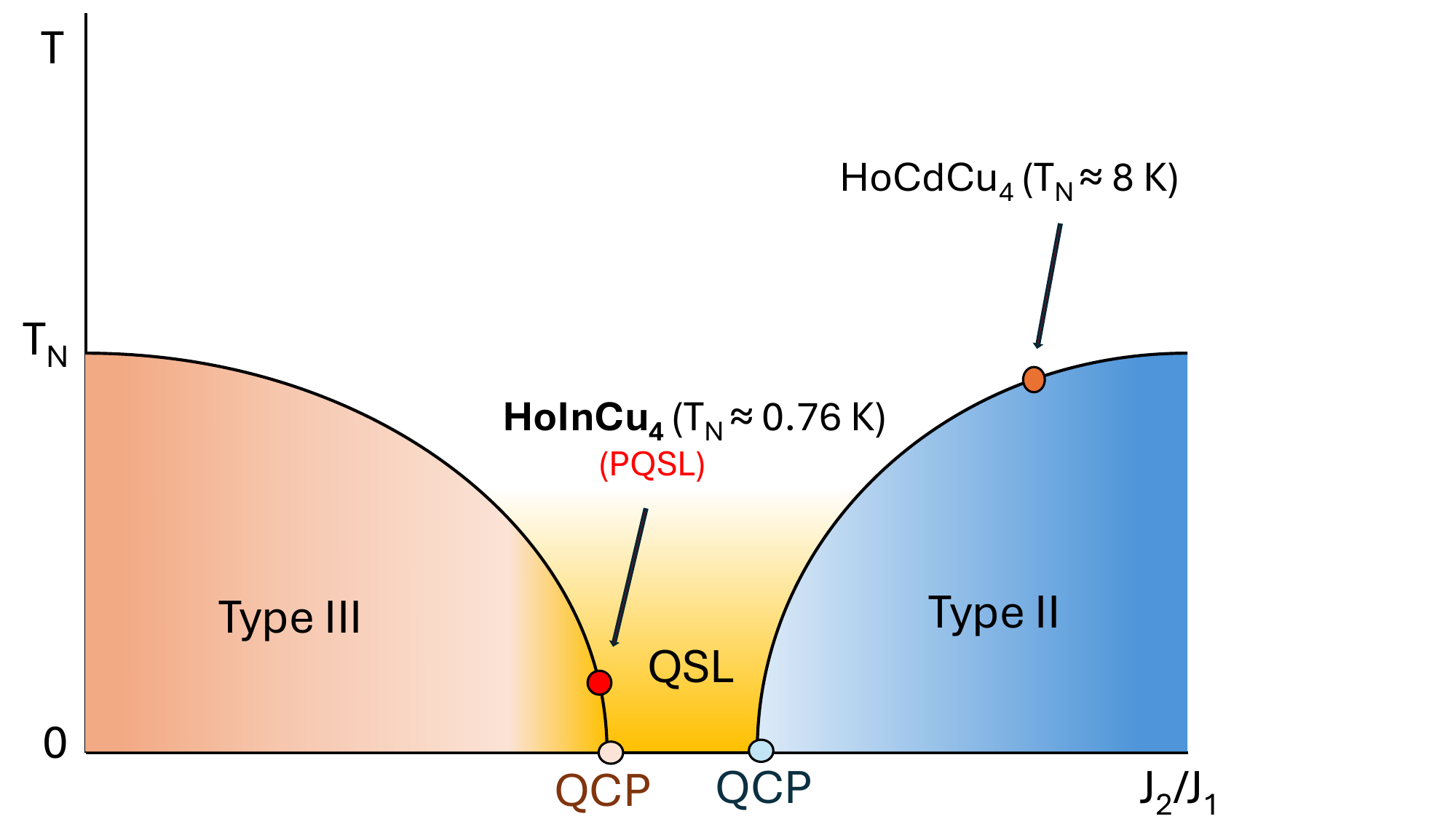}
\caption{\label{fig:phase-diagram}Phase diagram of magnetically
ordered systems vs.\ $J_2/J_1$ for an fcc-lattice. The depicted phases
include the different AFM ordering (type II and type III), quantum critical point (QCP), quantum spin liquid (QSL),
proximate quantum spin liquid (PQSL).
The two compounds studied here, HoCdCu$_4$ and HoInCu$_4$, are also
shown, along with the respective $T_\mathrm{N}$s.}  
\end{figure}
%%%%%%%%%%%%%%%%%%%%%%%%%%%%%%%%%%%%%%%%%%%%%%%%%%%%%%%%%%%%%%%%%%%%%%%%%%%%%%%

Summarizing the above results, Fig.~\ref{fig:phase-diagram} depicts
a schematic phase diagram where HoCdCu$_4$ is located well inside
the LRO region and away from the QCP or QSL state. In contrast,
HoInCu$_4$ is close to the QCP, but still within the LRO region, 
and in close proximity to a QSL state, as evidenced by the presence
of frustration-induced QSL-like features. 

By conducting detailed comparative analyses of the {\textmu}SR measurements,
we have characterized the magnetic ground state of the rare-earth intermetallic
compounds HoCdCu$_4$ and HoInCu$_4$, both featuring magnetic Ho$^{3+}$ 
moments arranged on a fcc lattice. Previous magnetization, heat capacity, and neutron diffraction studies
confirmed long-range antiferromagnetic order in both compounds, with
HoCdCu$_4$ showing no significant frustration, while HoInCu$_4$ displaying pronounced magnetic frustration. The investigation of both compounds helped us to
elucidate the nature of their ground states and the role of frustration
in stabilizing them. We identify the signature of frustration
in HoInCu$_4$ in the increase of $\lambda$ from a much higher temperature
compared to $T_\mathrm{N}$, in clear contrast to HoCdCu$_4$.
Moreover, in  HoInCu$_4$, only 30\% of the Ho$^{3+}$ moments
participate in the magnetic ordering below $T_\mathrm{N}$.
This is in contrast to HoCdCu$_4$, where all the Ho$^{3+}$ moments
are responsible for the ordering below $T_\mathrm{N}$. Interestingly,
$\lambda_f$, associated with the dynamic component of relaxation
(evident from LF measurements) shows a temperature independent behavior,
suggesting a persistent spin dynamics often seen in other QSL candidates.
Thus, below $T_\mathrm{N}$, there is a coexistence of static and
dynamic correlations, where the dynamic part corroborates the
experimental signatures of a QSL state.
Also, in HoInCu$_4$, the muon-spin relaxation rate exhibits a
power-law behavior, driven not only by thermal fluctuations but
also by quantum fluctuations. All the above mentioned properties
strongly suggest that HoInCu$_4$ exhibits a PQSL ground state.
This is rare in insulating systems and, to date, unobserved
in any frustrated metallic system. Our findings from {\textmu}SR
data are further supported by the similarities of recent inelastic
neutron results on HoInCu$_4$ with respect to the other insulating
PQSL candidates. Our study is expected to trigger further research
on HoInCu$_4$ and related systems. In particular, a QSL state beyond the
QCP could be achieved by applying magnetic field, hydrostatic, and
chemical pressure. Further, similar to HoInCu$_4$, other frustrated metallic
systems are expected to host proximate spin liquids.

\end{document}